\documentclass{article}
\usepackage{arxiv}
\usepackage[utf8]{inputenc} 
\usepackage[T1]{fontenc}    
\usepackage{url}            
\usepackage{booktabs}       
\usepackage{amsfonts}       
\usepackage{nicefrac}       
\usepackage{microtype}      
\usepackage{lipsum}		
\usepackage{graphicx}
\usepackage{natbib}
\usepackage{doi}
\usepackage{framed,multirow}
\usepackage{amssymb}
\usepackage{latexsym}
\usepackage{url}
\usepackage{xcolor}
\usepackage{threeparttable,booktabs}
\usepackage{hyperref}       

\usepackage{andrew_academic_paper}

\title{JParc: Joint cortical surface parcellation with registration}

\author{Jian~Li\txtsup{1,2,}\thanks{Corresponding author: jli112@mgh.harvard.edu} , Karthik~Gopinath\txtsup{1}, Brian~L.~Edlow\txtsup{1,2}, Adrian~V.~Dalca\txtsup{1,3,}\thanks{Co-senior authors with equal contribution} , Bruce~Fischl\txtsup{1,3,}\footnotemark[\value{footnote}]}
\affiliation{\txtsup{1} Athinoula A. Martinos Center for Biomedical Imaging, Department of Radiology, MGH \& HMS\\
\txtsup{2} Center for Neurotechnology and Neurorecovery, Department of Neurology, MGH \& HMS\\
\txtsup{3} Computer Science and Artificial Intelligence Laboratory, EECS, MIT}

\date{}

\renewcommand{\headeright}{}
\renewcommand{\undertitle}{}
\renewcommand{\shorttitle}{J. Li \textit{et al.} JParc: Joint cortical surface parcellation with registration}

\hypersetup{
pdftitle={JParc: Joint cortical surface parcellation with registration},
pdfsubject={eess.IV, cs.CV, q-bio.NC},
pdfauthor={Jian~Li, Karthik~Gopinath, Brian~L.~Edlow, Adrian~V.~Dalca, Bruce~Fischl},
}

\begin{document}
\maketitle

\begin{abstract}
  Cortical surface parcellation is a fundamental task in both basic neuroscience research and clinical applications, enabling more accurate mapping of brain regions. Model-based and learning-based approaches for automated parcellation alleviate the need for manual labeling. Despite the advancement in parcellation performance, learning-based methods shift away from registration and atlas propagation without exploring the reason for the improvement compared to traditional methods. In this study, we present JParc, a joint cortical registration and parcellation framework, that outperforms existing state-of-the-art parcellation methods. In rigorous experiments, we demonstrate that the enhanced performance of JParc is primarily attributable to accurate cortical registration and a learned parcellation atlas. By leveraging a shallow subnetwork to fine-tune the propagated atlas labels, JParc achieves a Dice score greater than 90\% on the Mindboggle dataset, using only basic geometric features (sulcal depth, curvature) that describe cortical folding patterns. The superior accuracy of JParc can significantly increase the statistical power in brain mapping studies as well as support applications in surgical planning and many other downstream neuroscientific and clinical tasks.\vspace*{2mm}
\end{abstract}


\section{Introduction}

Cortical surface parcellation, the division of the cerebral cortex into distinct regions, is essential in computational neuroscience research and a key step in many downstream neuroimaging applications \citep{Rademacher_1992_JCN_HumanCerebral,Destrieux_2010_N_AutomaticParcellation,Glasser_2016_N_MultimodalParcellation,Fischl_2018_N_MicrostructuralParcellation}. It is widely used in brain mapping studies, surgical planning, and longitudinal assessments of neurological and psychiatric disorders. An accurate cortical parcellation can greatly increase statistical power by reducing the multiple-comparison problem \citep{Eickhoff_2018_NRN_ImagingbasedParcellations,VanEssen_2012_N_HumanConnectome}.

Anatomical parcellation pipelines typically rely on brain surface geometry and topology, and subdivide cortex into regions of interests (ROIs) based on the anatomical boundaries of sulci and gyri \citep{Desikan_2006_N_AutomatedLabeling}. Using cortical geometric measures, such as curvature, derived from magnetic resonance imaging (MRI) data, manual cortical parcellation labeling of the cortical surfaces by neuroanatomy experts \citep{Klein_2012_FN_101Labeled} is a tedious task and requires large amounts of human effort.

Automated and semi-automated parcellation algorithms have significantly improved reproducibility and scalability compared to manual delineation. Compared to the registration-based traditional methods, modern deep learning methods predict cortical parcellation labels directly using the subject's cortical features. Despite the advancement in parcellation performance, learning-based studies shift away from registration and atlas propagation without exploring the reason for the improvement compared to traditional methods -- whether it stems from the enhanced network architectures or novel features, or from replacing the suboptimal registration accuracy inherent in traditional approaches.

In this work, we present a simple yet powerful framework for joint registration and cortical parcellation. Building on both traditional model-based methods and deep learning advances, our method, JParc, propagates labels from an atlas to the individual space based on cortical registration. We minimize inter-subject variance to achieve an accurate correspondence between subjects (registration), while jointly learning an optimal parcellation atlas that is used for label propagation. We enable the propagated labels to be fine-tuned using a shallow subnetwork. JParc achieves state-of-the-art performance using only basic input features (sulcal depth, curvature) that describe cortical folding patterns and a straightforward neural network design.

\section{Related Works}

Traditional model-based approaches often propagate manually curated atlas labels to the subject space using cortical registration, and generate individual parcellations using Bayesian segmentation \citep{Fischl_2004_CC_AutomaticallyParcellating}, probabilistic reasoning \citep{Desikan_2006_N_AutomatedLabeling,Fischl_2012_N_FreeSurfer}, clustering \citep{Craddock_2012_HBM_WholeBrain}, or classification \citep{Glasser_2016_N_MultimodalParcellation}. However, large individual variability in folding patterns poses a significant challenge in accurate cortical registration \citep{Li_2024_MIA_JOSAJoint,Fischl_2008_CC_CorticalFolding}, fundamentally limiting the performance of these strategies.

Deep learning techniques for cortical parcellation can provide performance and substantial speed improvement. A direct application of convolutional neural network (CNN) is often unsuccessful in parcellating the brain, because the cerebral cortical surface is a highly complex manifold in a non-Euclidean space \citep{Seong_2018_FN_GeometricConvolutional}. A variety of approaches expand on standard CNNs to improve performance. Some methods project cortical features, such as sulcal depth, curvature, onto the tangent plane at a given vertex and use a 2D CNN on the local feature patch to predict the label for that vertex \citep{Wu_2019_MICCAI-M2_IntrinsicPatchbased}. Spherical kernels enable convolutions to be performed directly on the sphere by designing a 1-ring kernel for convolution on different orders of icosahedrons \citep{Zhao_2019_IPMI_SphericalUnet,Zhao_2021_ITMI_SphericalDeformable} or applying a regular CNN on the parameterized 2D image \citep{Henschel_2020_BfdM2_ParameterSpace}. Novel spherical features, such as boundary maps \citep{Parvathaneni_2019_MICCAI-M2_CorticalSurface} or spherical harmonics \citep{Ha_2022_ITMI_SPHARMNetSpherical}, in addition to geometric features, can also enhance parcellation performance.

Graph convolutional networks (GCNs) enable learning intrinsic structural features directly in the non-Euclidean space \citep{Bronstein_2017_ISPM_GeometricDeep}, facilitating cortex parcellation using the original cortical surface. GCN was initially applied to parcellate Brodmann areas 44 and 45 \citep{Cucurull_2018_MIDL_ConvolutionalNeural}. Additional cortical features, such as spectral embedding \citep{Gopinath_2019_MIA_GraphConvolutions} or Cartesian and polar coordinates \citep{Zhang_2019_GLiMI_GeometricBrain}, have been added as additional inputs to improve the parcellation performance. Graph attention networks (GAT) have also been adapted to GCN for parcellation of both adult brains \citep{Eschenburg_2021_FN_LearningCortical} and fetal brains \citep{You_2024_FN_AutomaticCortical}.

\section{Methods}

We describe \textbf{J}oint registration-based cortical \textbf{Parc}ellation (JParc), a deep learning framework for cortical parcellation based on cortical registration. 

\subsection{Overview Model}
\label{sec:overview}

\begin{figure}[tb]
  \begin{center}\includegraphics[width=0.5\textwidth]{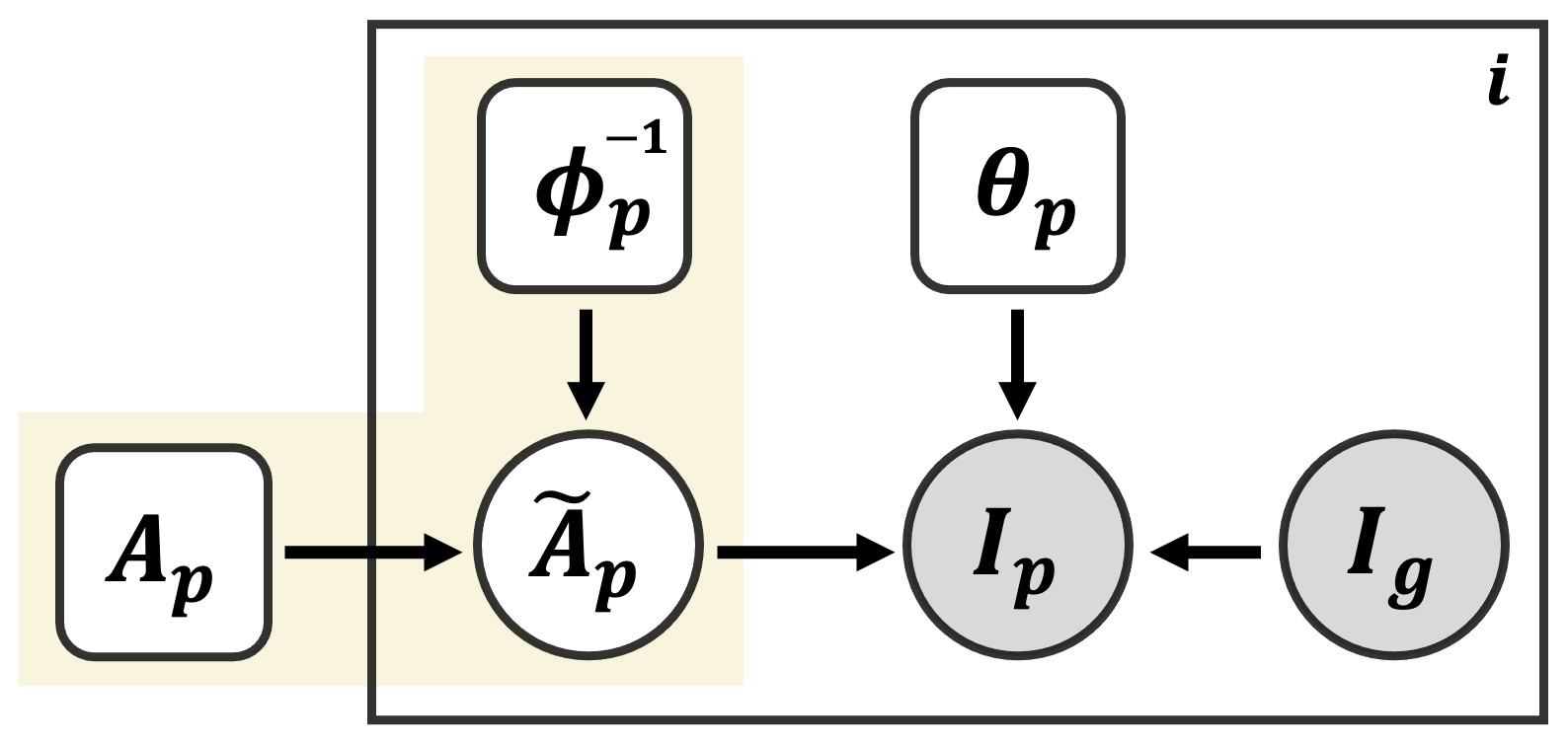}
    \caption{\textbf{Graphical representation of the probabilistic model.} Circles are random variables. Rounded squares indicate parameters. Shaded (gray) quantities are observations. The big plate represents replication over subjects ($i$). The subscript $p$ stands for parcellation and the subscript $g$ stands for geometry. $\A_{p}$ represents the parcellation atlas, $\tilde{\A_{p}}$ the deformed parcellation atlas (in the subject space) by the parcellation-specific deformation field $\phiB_{p}$. The operator $(\;\cdot\;)^{-1}$ indicates a warp inverse -- a warp from the atlas space to the subject space. $\I_{g}$ is the observed subject geometric image. $\I_{p}$ is the observed subject parcellation map. The yellow shaded area indicates the registration and $\thetaB_{p}$ is the parcellation prediction function (fine-tune).}
    \label{fig:model}
  \end{center}
\end{figure}
Fig.~\ref{fig:model} shows the graphical representation of the proposed model. Let $\A_{p}$ be the global parcellation atlas across the population. We propose a probabilistic model that generates the subject parcellation map $\I_{p}$ by first warping the parcellation atlas $\A_p$ using a parcellation-specific deformation $\phiB_{p}$. Then the warped parcellation atlas $\tilde{\A}_{p}$ (in the subject space) is fine-tuned by the parcellation prediction function $\thetaB_{p}$ in combination with the subject geometry $\I_{g}$. All variables except for the global atlas are subject-specific, hence we omit $i$ from the following derivation.

Specifically, $\tilde{\A}_{p}$ is a noisy observation of the warped atlas
\begin{equation}
  \label{eq:likelihood}
  \mathbb{P}(\tilde{\A}_{p} | \phiB_{p}^{-1}; \A_{p}) = \gauss(\tilde{\A}_{p}; \phiB_{p}^{-1}\circ \A_{p}, \sigma^{2}\mathbb{I}),
\end{equation}
where $\gauss(\;\cdot\;; \muB, \SigmaB)$ is the multivariate Gaussian distribution with mean $\muB$ and covariance $\SigmaB$, $\sigma$ represents the variance of additive noise, and $\mathbb{I}$ is the identity matrix.

We impose the following prior for the deformation $\phiB_{p}^{-1}$
\begin{equation}
  \label{eq:prior}
  \mathbb{P}(\phiB_{p}^{-1})\; \sim\; \text{exp}\{-\lambda \norm{\nabla \u_{p}}^{2}\},
\end{equation}
where $\u_{p}$ is the spatial displacement for $\phiB_{p}^{-1} = Id + \u_{p}$, $\nabla \u_{p}$ is its spatial gradient.

Let $K=\{1, 2, \dots, n\}$ be the set of $n$ discrete parcellation labels. The parcellation map $\hat{\I}_{p}$ has a categorical probabilistic distribution
\begin{equation}
  \label{eq:categotrical}
  \mathbb{P}(\I_{p}=k | \thetaB_{p}; \tilde{\A}_{p}, \I_{g}) = \thetaB_{p}(\tilde{\A}_{p}, \I_{g}) = y_{k}, \; k\in K,
\end{equation}
where $y_{k}$ represents the probability of label $k$, and $\sum_{k=1}^{K}y_{k}=1$.

\subsection{Learning}
\label{sec:learning}

We learn the population atlas $\A_{p}$ and the deformation field $\phiB_{p}^{-1}$ by minimizing the negative log likelihood of $\phiB_{p}^{-1}$
\begin{equation}
  \label{eq:min-log}
  \begin{alignedat}{4}
    \mathcal{L}(\phiB_{p}^{-1} | \tilde{\A}_{p}; \A_{p}) &= && -\log \mathbb{P}(\phiB_{p}^{-1} | \tilde{\A}_{p}; \A_{p}) = -\log \mathbb{P}(\tilde{\A}_{p} | \phiB_{p}^{-1}; \A_{p}) - \log \mathbb{P}(\phiB_{p}^{-1})\\
    &= &&\quad \frac{1}{2\sigma^{2}} \norm{\tilde{\A}_{p}-\phiB_{p}^{-1}\circ \A_{p}}^{2} + \lambda\norm{\nabla\u_{p}}^{2} + \text{const.}
  \end{alignedat}
\end{equation}

To learn the prediction function $\thetaB_{p}$, we minimize the soft Dice loss
\begin{equation}
  \label{eq:min-dice}
  \mathcal{L}(\thetaB_{p} | \I_{p}; \tilde{\A}_{p}, \I_{g}) = 1 - Dice(\thetaB_{p}(\tilde{\A}_{p}, \I_{g}), \I_{p}) = 1 - \frac{2\sum_{k\in K}y_{k}\delta(\I_{p}=k)}{\sum_{k\in K}y_{k}^{2} + \sum_{k\in K}{\delta(\I_{p}=k)}^{2}},
\end{equation}
where $\delta(\cdot)$ is the Derac delta function.

\subsection{Network Architecture}
\label{sec:network}

\begin{figure}[tb]
  \begin{center}\includegraphics[width=0.8\textwidth]{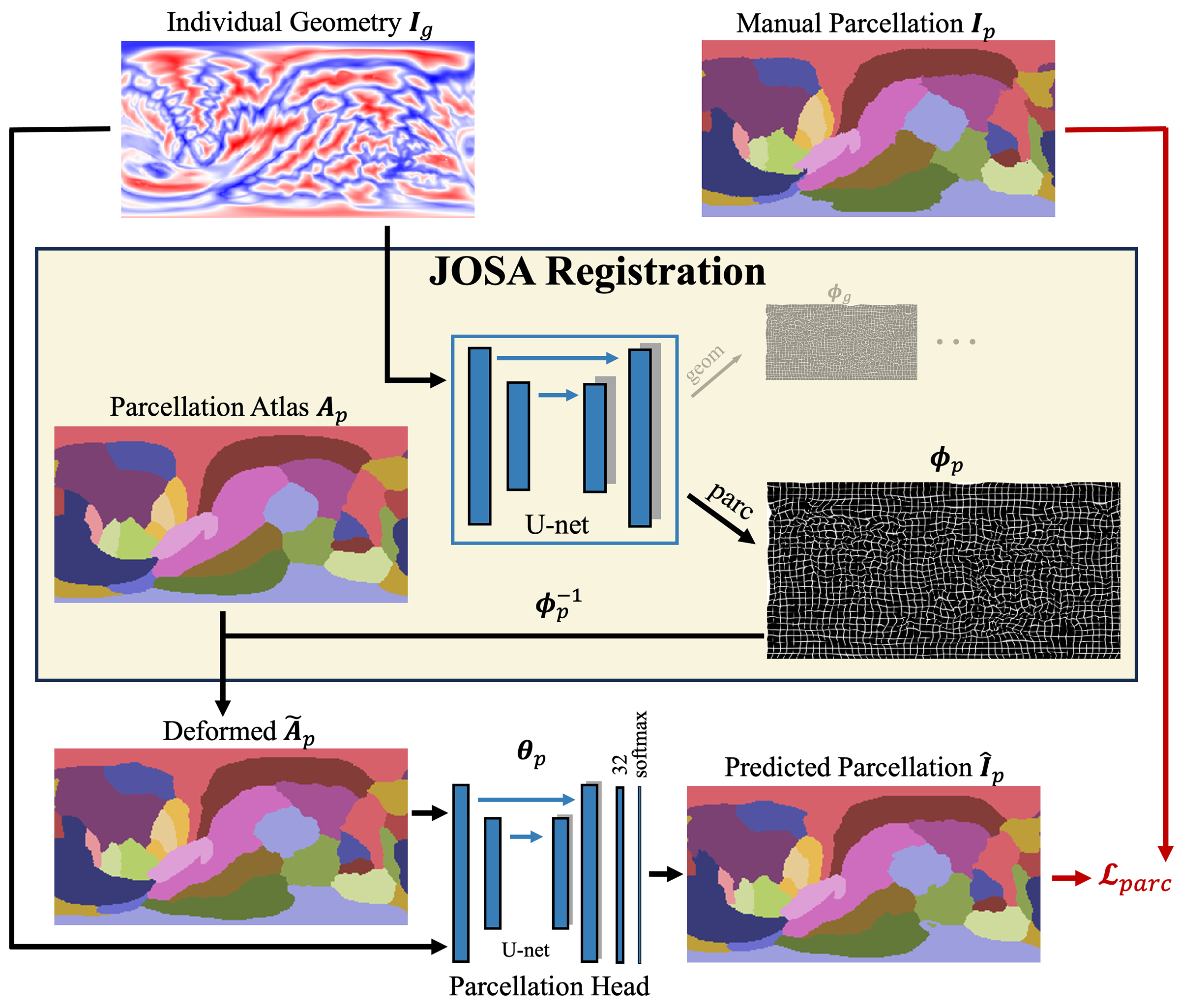}
    \caption{\textbf{Overview of JParc network architecture.} The JOSA registration module (yellow box) takes individual geometric features $\I_{g}$ and propagates the parcellation atlas $\A_{p}$ to the individual space through the parcellation-specific deformation. The deformed parcellation atlas $\tilde{\A}_{p}$, concatenated with the individual geometric features $\I_{g}$, are input to a parcellation head $\thetaB_{p}$ to generate the final prediction on the parcellation $\hat{\I}_{p}$. The predicted parcellation $\hat{\I}_{p}$ is compared to the manual parcellation $\I_{p}$ for loss evaluation.}
    \label{fig:network}
  \end{center}
\end{figure}

As illustrated in Fig.~\ref{fig:network}, JParc takes the geometric features and predicts the cortical parcellation map for each individual subject. All data have been projected onto the parameterized space for training and inference \citep{Li_2024_MIA_JOSAJoint,Henschel_2020_BfdM2_ParameterSpace,Cheng_2020_N_CorticalSurface}. The only input to the networks is the vector of geometric features $\I_{g}$ describing the cortical folding patterns of the input brain. The mean curvature map is shown for illustration purposes (top-left image) in Fig.~\ref{fig:network}.

We use JOSA \citep{Li_2024_MIA_JOSAJoint} registration (shaded in yellow) to 1) learn an optimal parcellation atlas $\A_{p}$ (as part of the network parameters); 2) produce an optimal deformation $\phiB_{g}$ for geometric registration; and 3) an optimal deformation $\phiB_{p}$ for alignment of parcellation maps. Although the geometric branch of JOSA ($\phiB_{g}$) is not further used in this JParc framework, we highlight the importance of the two separated but also closely-related registration paths, $\phiB_{g}$ vs $\phiB_{p}$. The two registration paths not only capture the large inter-subject variance but also provide the flexibility to warp each modality in a slightly different way to achieve optimal alignment in each without compromising the other.
Using the diffeomorphic property of the deformation field, we warp the parcellation atlas $\A_{p}$ using $\phiB_{p}^{-1}$. The deformed parcellation atlas $\tilde{\A}_{p}$, concatenated with the original geometric features $\I_{g}$, is passed through a small U-net followed by a convoluational layer with spherical paddings and a final softmax layer (together denoted as the parcellation head $\theta_{p}$) to generate the predicted the individual parcellation map $\hat{\I}_{p}$. Finally, we compared the predicted parcellation $\hat{\I}_{p}$ with the manual parcellation map $\I_{p}$ (the ``ground truth'') for loss evaluation using the Dice metric ($\mathcal{L}_{parc}$).

\subsection{Implementation Details}
\label{sec:data-imp}

We implemented JOSA as described in \citet{Li_2024_MIA_JOSAJoint}, where the core architecture is based on VoxelMorph unsupervised framework \citep{Dalca_2019_MIA_UnsupervisedLearning,Balakrishnan_2019_ITMI_VoxelMorphLearning}. We used 5 encoder layers and 7 decoder layers with a flat 128 filters (i.e., channels or features) for each layer in the JOSA U-net. We mapped all data, including the inputs, the results from intermediate layers, and the final outputs, to the parameterized space with a spherical padding of 16 pixels on each side, where a 180-degree circular shift and a reflection was performed on the top and bottom edges of the images and a circular rolling was performed to the left and right sides of the images to ensure consistent convolution across the boundaries \citep{Li_2024_MIA_JOSAJoint,Li_2023_MIDL_JointCortical,Henschel_2020_BfdM2_ParameterSpace,Cheng_2020_N_CorticalSurface}. The input geometric features include the sulcal depth (\verb|sulc|), mean curvature (\verb|curv|), and the FreeSurfer curvature map from the inflated surface (\verb|inflated.H|).

We used a small U-net, one additional convolutional layer, and a final softmax layer for the parcellation head. We chose a U-net with a 3-layer encoder ([64, 128, 256] filters) and a symmetric decoder empirically to achieve good parcellation performance but without substantial overfitting. The last convolutional layer consists of 32 channels that correspond to 32 ROIs defined in the Desikan-Killiany-Tourville (DKT) parcellation protocol \citep{Klein_2012_FN_101Labeled,Desikan_2006_N_AutomatedLabeling}. In the loss evaluation, we weighted the Dice score $\mathcal{L}_{parc}$ by $\sin{(\theta)}$ to account for the metric distortion during parameterization, where $\theta$ represents the elevation/polar angle in the spherical coordinate system \citep{Li_2024_MIA_JOSAJoint,Cheng_2020_N_CorticalSurface}.

During training, we used a batch size of 8 and the Adam optimizer \citep{Kingma_2014_AdamMethod} with an initial learning rate of $10^{-4}$. The learning rate linearly decayed to $10^{-5}$ at the 500\txtsup{th} epoch and then further decreased by a factor of 0.99 as needed when the validation loss did not improve for 100 epochs. Due to the limited number of subjects ($N=100$) in the dataset (see Data section below), we randomly split the subjects into 5 folds and employed a 5-fold cross-validation. Each time, we held 1 fold ($N=20$) for testing and combined the remaining 4 folds ($N=80$) for training ($N=70$) and validation ($N=10$). To ensure data independence across different folds, we pre-trained the JOSA module using the same set of subjects as those used in the JParc training, separately for each split. We then fixed the network parameters in the JOSA module (including the parcellation atlas $\A_{p}$) and learned parameters in the parcellation head only during JParc training.

We used Tensorflow \citep{Abadi_2016_TensorFlowLargeScale}, with a Keras front-end \citep{Chollet_2018_ASCL_KerasPython} and the Neurite package \citep{Dalca_2018_MICCAI_UnsupervisedLearning} as the software environment. All training was conducted in our in-house high-performance computing cluster. Typical resources allocated for each job are 4 CPU cores, 64 GB of CPU memory, and an Nvidia GPU (A100 or RTX8000 or RTX6000).

\section{Experiments and Results}
\label{sec:exp-res}

\subsection{Data}
\label{sec:data}

We used Mindboggle \citep{Klein_2012_FN_101Labeled}, the largest publicly available dataset in which manual labels were provided for each subject. Manual parcellation was performed using the Desikan-Killiany-Tourville (DKT) protocol, a modified Desikan-Killiany protocol \citep{Desikan_2006_N_AutomatedLabeling} where labels for banks of superior temporal sulcus, temporal pole, and frontal pole were removed due to the lack of clear geometric landmarks. The resulting parcellation maps have 32 ROIs (including one for midwall/unknown), consistent across all subjects. In total, there are 100 subjects available with complete FreeSurfer recon outputs (v5.1) and manual labels.

\subsection{Baseline Comparison}
\label{sec:comparison}

We compared JParc with the following state-of-the-art approaches that also used the Mindboggle as the benchmark dataset:
\begin{itemize}
\item FreeSurfer \citep{Fischl_2012_N_FreeSurfer}
\item Euclidean Random Forest (ERF) \citep{Lombaert_2015_MICaCI-M2_SpectralForests}
\item Spectral Random Forest (SRF) \citep{Lombaert_2015_MICaCI-M2_SpectralForests}
\item Deep Brain Parcellation Network (DBPN) \citep{Zhang_2019_GLiMI_GeometricBrain}
\item Unstructured Grid Spherical CNN (UGSCNN) \citep{Parvathaneni_2019_MICCAI-M2_CorticalSurface}
\item Spectral Graph Convolutional Network (GCN) \citep{Gopinath_2019_MIA_GraphConvolutions}
\item SPherical HARMonics-based CNN (SPHARM-Net) \citep{Ha_2022_ITMI_SPHARMNetSpherical}
\item nnU-Net \citep{Isensee_2021_NM_NnUNetSelfconfiguring}
\end{itemize}

For previous studies we have access to either their publicly available models or executables (FreeSurfer, ERF, SRF, GCN), we performed a single forward pass inference or ran the executable to generate the predicted parcellation for each subject in the Mindboggle dataset. We computed the evaluation metrics using the predicted and manual labels and provided both qualitative and quantitative comparisons below. For other approaches in the baseline list where the evaluation was performed and cross-validated on Mindboggle, we listed the available evaluation metrics based on the results reported directly in their papers. For nnU-Net, we trained the model and evaluated the performance ourselves using their recommended/default configurations and hyperparameters according to the public repository and instructions (\url{https://github.com/MIC-DKFZ/nnUNet}).

\subsection{Evaluation Metrics}
\label{sec:evaluation-metrics}

We measured the parcellation performance using the Dice score:
\begin{equation*}
  \text{Dice}\; (L_{p}, L_{m}) = \frac{2|L_{p}\bigcap L_{m}|}{|L_{p}|+|L_{m}|}
\end{equation*}
and point-wise accuracy:
\begin{equation*}
  \text{Accuracy}\; (L_{p}, L_{m}) = \frac{|L_{p}\bigcap L_{m}|}{|L_{m}|}
\end{equation*}
where $L_{p}$ and $L_{m}$ are the predicted label and manual label, respectively and $| \cdot |$ is the cardinality. We computed both metrics for each ROI and each subject. Median and interquartile range of the Dice scores across subjects are shown in Fig.~\ref{fig:dice} and the average Dice and accuracy across ROIs are reported in Table~\ref{tab:comparison} as an overall comparison.

\subsection{Statistics}
\label{sec:stats}

For FreeSurfer, ERF, SRF, GCN, and nnU-Net where we have access to the parcellation results for each individual subject, we performed the pair-wise (subject) Wilcoxon signed-rank test between JParc and each baseline method. We tested the difference in the mean Dice and mean accuracy over all ROIs and applied a Bonferroni correction \citep{Bonferroni_1936_TeoriaStatistica} to the resulting p-values (Table~\ref{tab:comparison}). We further tested the difference for each ROI between JParc and all other baseline methods and applied the Benjamini-Hochberg method \citep{Benjamini_1995_JRSSSBSM_ControllingFalse} to control the false discovery rate (FDR) across all ROIs and method pairs (Fig.~\ref{fig:dice}).

\subsection{Ablation Studies}
\label{sec:ablation}

We performed two ablation studies. First, we removed the parcellation head and used the propagated atlas labels (the direct output from the JOSA registration module, i.e., the bottom left image ``Deformed $\A_{p}$ in Fig.~\ref{fig:network}) as the final parcellation prediction for each subject. We evaluated the performance using Dice and accuracy, identical to the evaluation used in the main experiments.

The second ablation experiment was identical to the first ablation experiment, except that we used the geometric deformation field ($\phiB_{g}$ in the semi-transparent geometric path in Fig.~\ref{fig:network}) to warp the parcellation atlas $\A_{p}$ instead of using the parcellation deformation  ($\phiB_{p}$).

\subsection{Results}
\label{sec:result}

\subsubsection{Qualitative Comparison}
\label{sec:res:qual}

\begin{figure}[tb]
  \centering
  \includegraphics[width=1\textwidth]{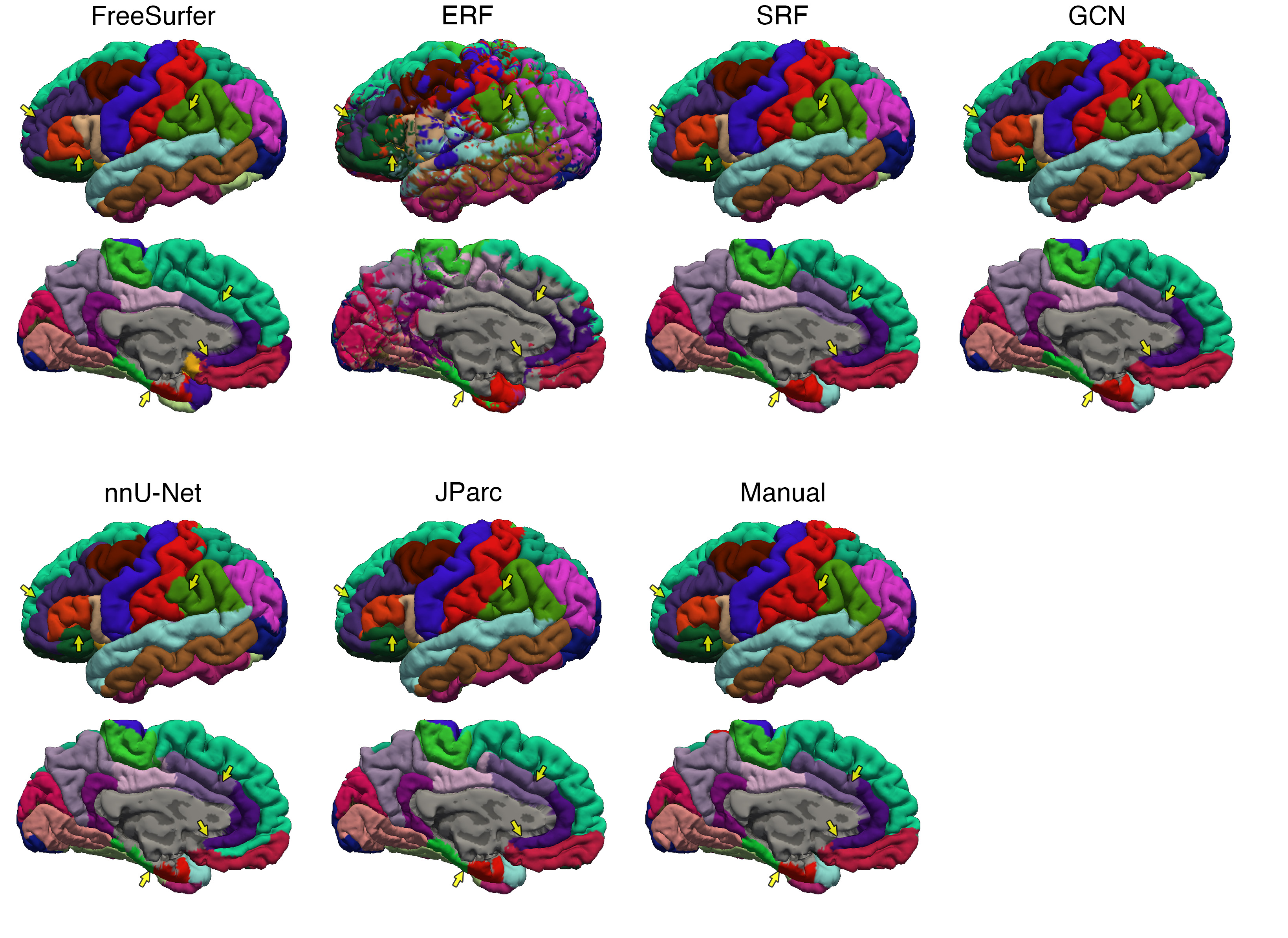}
  \caption{\textbf{Visual comparison of parcellation maps for an exemplar subject.} Lateral surface view is shown on top and medial surface view is shown on the bottom for each method. Yellow arrows indicate regions where JParc substantially outperforms other baseline methods in comparison to the manual parcellation map.}
  \label{fig:qual}
\end{figure}

Fig.~\ref{fig:qual} shows a qualitative comparison between JParc and baseline methods for an exemplar subject. For each method, we show the lateral surface view on top and the medial surface view below for the same subject. Method names are indicated on top of each column. We overlaid the parcellation maps with the standard ROI colors used in FreeSurfer \citep{Fischl_2012_N_FreeSurfer}. JParc achieved a substantially higher similarity with the manual parcellations than the baseline methods. This is particularly evident in regions involving higher cognitive functions, such as paropercularis, parstriangularis, parsorbitalis, superior frontal gyrus, inferior parietal lobule, caudal/rostal anterior cingulate, and parahippocampal/entorhinal cortex, as indicated by the yellow arrows. This result is expected as higher-order cognitive areas exhibit greater inter-subject variance in both geometry and function than primary regions \citep{Fischl_2008_CC_CorticalFolding,Frost_2012_N_MeasuringStructural}.

\begin{table}[tb]
  \centering
  \begin{threeparttable}
  \caption{Parcellation performance comparison. The overall Dice and its standard deviation (std) and the counterpart for accuracy are shown in the second and third columns, respectively. The Bonferroni corrected p-values are shown inside parentheses where the data are available.}
  \label{tab:comparison}
  \begin{tabular}{lll}
    \toprule
    Methods  & (Dice $\pm$ std)\% (p-value) & (Accuracy $\pm$ std)\% (p-value)\\
    \midrule
    FreeSurfer & $84.01 \pm 1.68$ ($1.0\times 10^{-17}$) & $84.03 \pm 1.85$ ($1.0\times 10^{-17}$)\\
    ERF & $45.87\pm 8.74$ ($9.7\times 10^{-18}$) & $49.26\pm 8.32$ ($9.7\times 10^{-18}$)\\
    SRF & $79.89 \pm 2.62$ ($1.0\times 10^{-17}$) & $81.94 \pm 2.54$ ($1.4\times 10^{-17}$)\\
    DBPN\tnote{$\dagger$} & $84.60 \pm 3.40$ & -\\
    UGSCNN\tnote{$\dagger$} & $86.28 \pm 2.54$ & -\\
    GCN & $86.61 \pm 2.45$ ($2.6\times 10^{-16}$) & $88.08 \pm 2.47$ ($1.9\times 10^{-14}$)\\
    SPHARM-Net\tnote{$\dagger$} & $88.64\pm 1.83$ & -\\
    nnU-Net & $88.17 \pm 1.77$ ($1.3\times 10^{-17}$) & $88.72 \pm 1.95$ ($1.6\times 10^{-17}$)\\
    \midrule
    \textbf{JParc} & \bm{$90.20 \pm 1.77$} & \bm{$90.96 \pm 1.80$}\\
    \midrule
    JOSA-Geom & $86.88 \pm 1.76$ & $88.37 \pm 1.60$\\
    JOSA-Parc & $88.38 \pm 1.75$ & $89.63 \pm 1.51$\\
    \bottomrule
  \end{tabular}
  \begin{tablenotes}[flushleft]\footnotesize
  \item[$\dagger$]Parcellation performed on lower resolution surfaces (e.g. 5th or 6th order icosahedron).
  \end{tablenotes}
\end{threeparttable}
\end{table}

\subsubsection{Quantitative Comparison}
\label{sec:res:quant}

Table~\ref{tab:comparison} summarizes the overall Dice score and its standard deviation (std) for JParc and baseline methods in the second column as well as the mean accuracy metric and its std for methods where available in the third column. JParc yields over $90\%$ of Dice and accuracy that significantly outperforms all other baseline approaches as indicated by the p-values.

\begin{figure}[tb]
  \centering
  \includegraphics[width=1\textwidth]{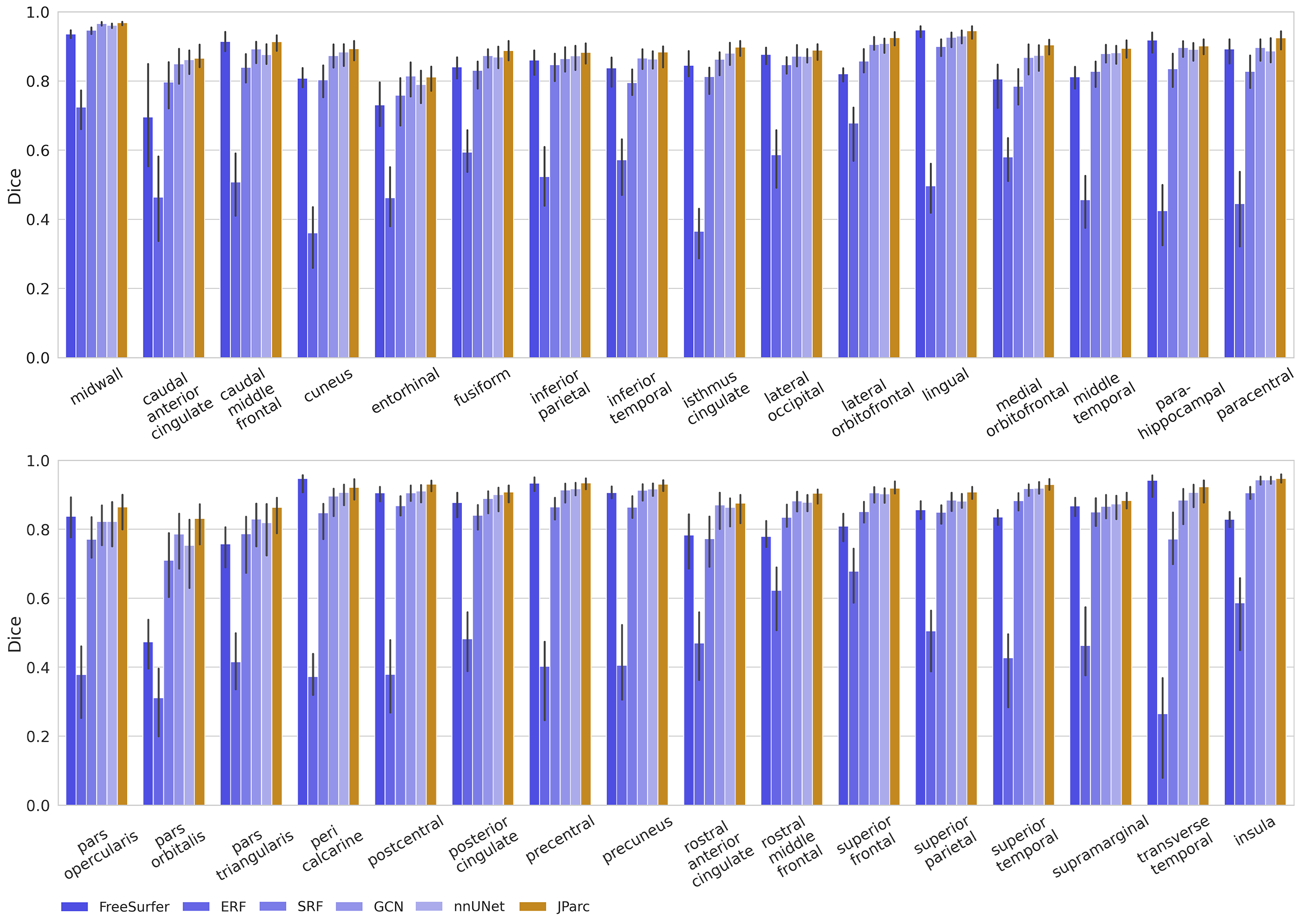}
  \caption{\textbf{Bar plots of Dice coefficients for each of the 32 ROIs.} The ROIs' names are shown along the x-axis and the Dice score is shown on the y-axis. The box represents the median Dice and the black vertical lines through the center of each box indicate the interquartile range (middle 50\%). Baseline methods are shown in various blue shades and JParc is shown in orange.}
  \label{fig:dice}
\end{figure}

Fig.~\ref{fig:dice} breaks down the Dice measures into each of the 32 ROIs that were defined in the DKT protocol. The results confirm our visual observations that JParc Dice is significantly higher than the baseline methods in most of the ROIs, and differs by a large margin in regions such as parsorbitalis, caudal anterior cingulate. Wilcoxon signed-rank tests show significant difference with FDR corrected $p<10^{-3}$ for 120 out of 160 pairs (5 method pairs x 32 ROIs) of tests, $p<10^{-2}$ for 137 pairs, and $p<0.05$ for \textbf{all} pairs except for following 12 pairs: the JParc-FreeSurfer pairs in caudal middle frontal, lingual, para-hippocampal, pericalcarine, precentral, and transverse temporal regions, the JParc-GCN pairs in midwall, entorhinal, para-hippocampal, and the rostral anterior cingulate, and the JParc-nnUNet pairs in posterior cingulate and rostral anterior cingulate. The parcellation results from JParc are as robust as, if not more robust than, the baseline methods, as indicated by the interquartile range.

\subsubsection{Unsatisfactory Cases}
\label{sec:res:bad}

\begin{figure}[tb]
  \centering
  \includegraphics[width=.8\textwidth]{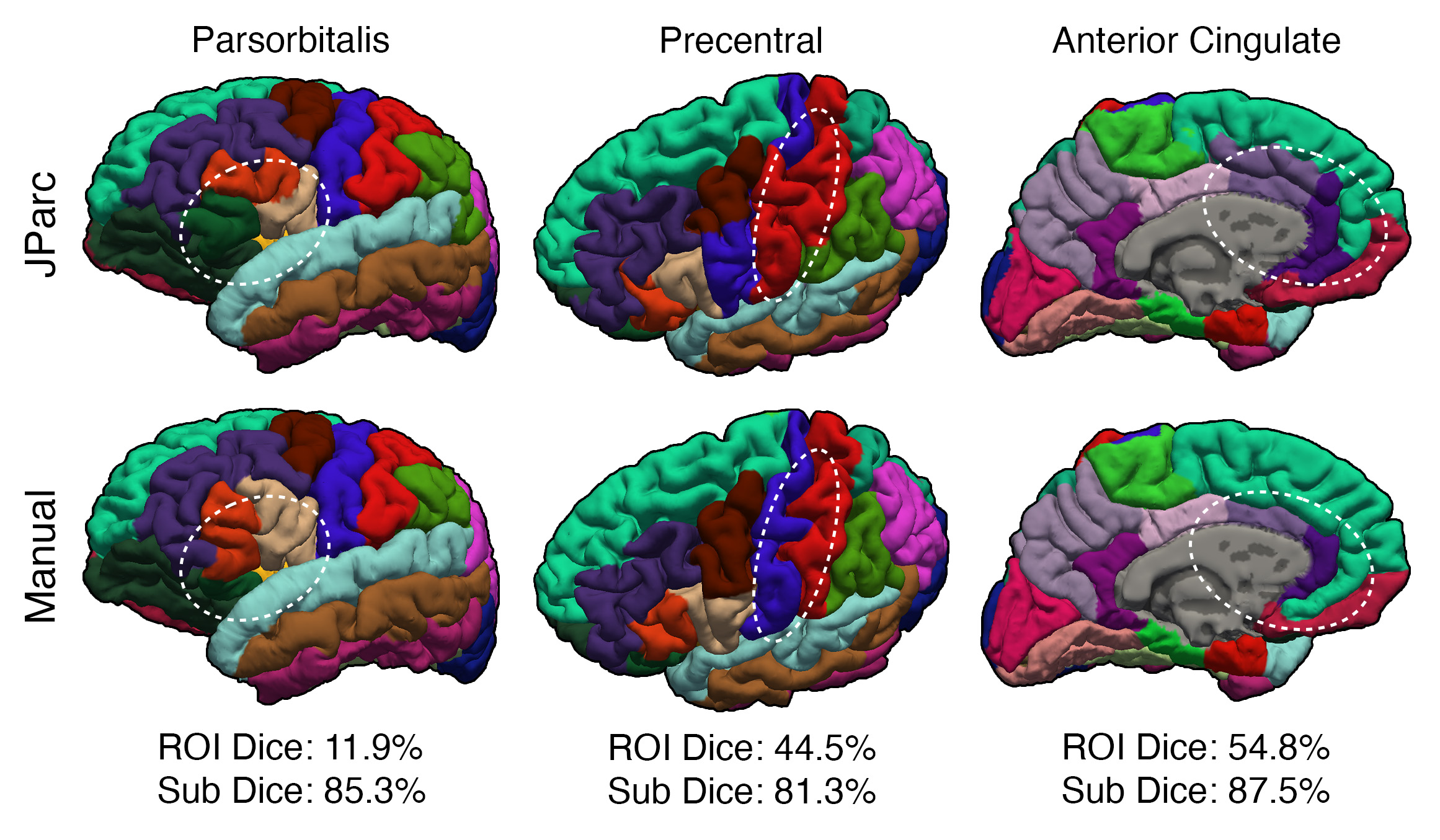}
  \caption{\textbf{Examples of unsatisfactory parcellation for some ROIs.} Cases for parsorbitalis, precentral, and anterior cingulate are shown in the left, middle, and right columns, respectively. The top row shows the JParc result and the bottom row shows the manual parcellation map. The Dice for the corresponding ROI and the overall ROI for that subject are shown below the images in each column.}
  \label{fig:bad}
\end{figure}

Fig.~\ref{fig:bad} illustrates some of the ``bad'' parcellation examples for parsorbitalis, precentral, and anterior cingulate in each of three columns, respectively. Compared with the manual labels on the bottom row, JParc substantially shifted the labels across sulci/gyri (parsorbitalis case), or labeled significantly less (precentral case) or more (anterior cingulate case) vertices within or around the ROI. Parcellation performance in regions other than that ``problematic'' ROI and its neighorhood remains reasonably high accuracy, highlighting the robustness of the JParc pipeline.

\subsubsection{Ablation Studies}
\label{sec:res:ablation}

When using a pure registration-based approach with dedicated parcellation deformation, Table~\ref{tab:comparison}, JOSA-Parc yields $88.38\%$ Dice and $89.63\%$ accuracy, which is, although suboptimal compared to JParc, still very comparable to top performers among the state-of-the-art methods. This result highlights the importance of inter-subject cortical registration in the ability to generate accurate cortical parcellation. In the second ablation experiment where we used the geometric deformation field ($\phiB_{j}\circ \phiB_{g}$) to warp the parcellation atlas, we observe a further slightly decreased parcellation performance with $86.88\%$ Dice and $88.37\%$ accuracy as shown in Table~\ref{tab:comparison} JOSA-Geom.

\section{Conclusion and Discussion}
\label{sec:discussion}
Cortical surface parcellation is important not only in basic neuroscience research, particularly in brain mapping and the study of neurological and psychiatric disorders, but also as a key tool in surgical planning and other clinical applications \citep{Eickhoff_2018_NRN_ImagingbasedParcellations}. Historically, cortical parcellation relied predominantly on manual labeling by neuroanatomy experts \citep{Klein_2012_FN_101Labeled}. With advances in computational neuroimaging, numerous automated parcellation methods have been developed to reduce manual effort and facilitate scalability for large datasets \citep{Fischl_2004_CC_AutomaticallyParcellating,Desikan_2006_N_AutomatedLabeling,Fischl_2012_N_FreeSurfer,Craddock_2012_HBM_WholeBrain,Glasser_2016_N_MultimodalParcellation}. Recent breakthroughs in deep learning techniques have further enhanced the performance of cortical parcellation, significantly accelerating its speed at inference \citep{Wu_2019_MICCAI-M2_IntrinsicPatchbased,Zhao_2021_ITMI_SphericalDeformable,Zhao_2021_MICaCAI-M2_DeepNetwork,Henschel_2020_BfdM2_ParameterSpace,Parvathaneni_2019_MICCAI-M2_CorticalSurface,Ha_2022_ITMI_SPHARMNetSpherical,Cucurull_2018_MIDL_ConvolutionalNeural,Gopinath_2019_MIA_GraphConvolutions,Gopinath_2023_MIA_LearningJoint,Eschenburg_2021_FN_LearningCortical,Li_2022_CiBaM_AnatomicallyConstrained,You_2024_FN_AutomaticCortical}.

In this study, we introduce JParc, a joint cortical registration and parcellation framework that achieves superior parcellation performance compared to state-of-the-art methods. Unlike approaches that rely on complex network architectures or the extraction of various novel features from MR images, JParc utilizes our recently developed cortical registration method, JOSA \citep{Li_2024_MIA_JOSAJoint}, to propagate atlas labels into individual subject spaces, serving as the foundation for parcellation. Our results demonstrate that, even in the absence of fine-tuning with a dedicated parcellation head (Fig.~\ref{fig:network}), we can still achieve an accurate parcellation performance, contingent on an accurate cortical registration, as evidenced by our ablation studies.

We believe the large inter-subject variance in cortical folding patterns is the primary factor limiting the parcellation performance. Although cortical folding patterns are closely related to cortical parcellation, as manual labeling is primarily guided by these anatomical landmarks \citep{Desikan_2006_N_AutomatedLabeling,Destrieux_2010_N_AutomaticParcellation,Klein_2012_FN_101Labeled}, this relationship is not a straightforward one-to-one point-wise correspondence due to substantial inter-subject variability in cortical folding patterns. Indeed, prior work has demonstrated that optimal geometric alignment does not necessarily produce optimal parcellation \citep{Parvathaneni_2019_MICCAI-M2_CorticalSurface,Fischl_2008_CC_CorticalFolding} or functional correspondence \citep{Li_2023_MIDL_JointCortical,Li_2024_MIA_JOSAJoint}. In this study, it is the parcellation-dedicated deformation ($\phiB_{p}$) and the optimally learned atlases ($\A_{p}$) inherited from the JOSA registration that place us in a significantly better position to establish accurate inter-subject correspondence compared to traditional parcellation methods, thereby yielding enhanced parcellation performance in JParc.

\begin{figure}[tb]
  \centering
  \includegraphics[width=.6\textwidth]{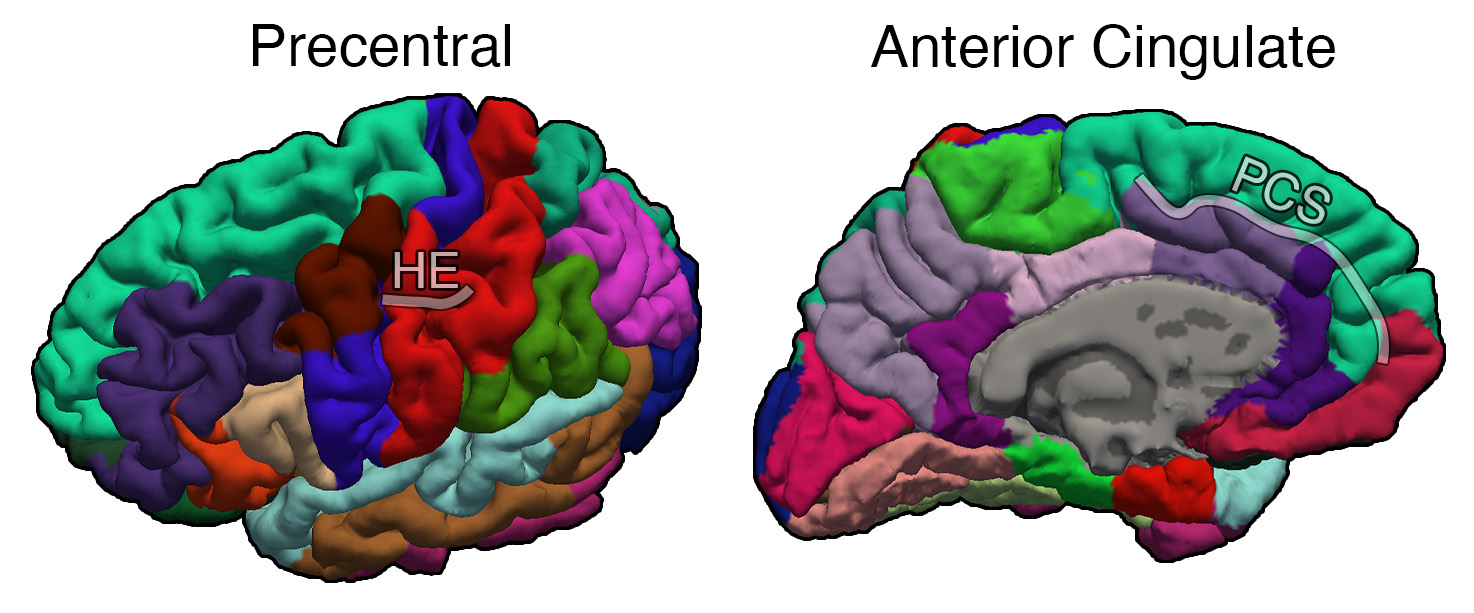}
  \caption{\textbf{Two of the unsatisfactory parcellation examples.} The left column shows an example for the precentral ROI with the horizontal extension of the interior precentral sulcus (HE) present and the right column shows an example for the anterior cingulate with the paracingulate sulcus (PCS) marked, respectively.}
  \label{fig:var}
\end{figure}

\subsection{Inter-subject Variability}
\label{sec:discuss:var}

To better illustrate the impact of inter-subject variability on parcellation, we re-examined the failure cases previously shown in Fig.~\ref{fig:bad} and re-drew them in Fig.~\ref{fig:var}. In the precentral case, the JParc label was disrupted by the discontinuity of the precentral gyrus, which results from an unusually long posterior extension of the horizontal branch of the inferior precentral sulcus (HE) \citep{Germann_2005_JCN_PrecentralSulcal}. Although this morphological variation is not commonly observed across the population, the precentral gyrus can still be robustly labeled by hand when the extended HE is present. However, this neuroanatomical prior (the precentral and postcentral gyri should run from the midline to the lateral fissure in parallel) was not learned by JParc as no subject in the training set exhibited a similar HE pattern, indicating that a significantly expanded training set could ameliorate this type of issue. 

The second case involving anterior cingulate parcellation is more debatable, Fig.~\ref{fig:var} right panel. Two major morphological variations of the anterior cingulate are commonly observed: one with a single gyrus and another with two parallel gyri \citep{Fornito_2008_HBM_VariabilityParacingulate,Lahutsina_2022_J_MorphologyAnterior}. The presence of the paracingulate sulcus (PCS) typically determines the specific pattern. In this outlier case, the PCS may not be sufficiently deep or continuous through the medial frontal cortex. However, it remains within the realm of subjective judgment to assert the presence of the PCS. Many cases within the dataset (and the broader population) exhibit a more distinct PCS, allowing both gyri to be confidently labeled as part of the anterior cingulate.

\subsection{Potential Biases in Evaluation}
\label{sec:discuss:bias}

Although JParc achieved higher Dice scores than the state-of-the-art methods, the top performers differ by only a few percentage points. As improving parcellation performance becomes increasingly challenging when training and validation are conducted solely within the Mindboggle dataset, we would like to highlight several considerations that may introduce potential bias in performance evaluation.

The use of the Dice coefficient in evaluation tends to favor larger regions, as it is more tolerant of errors made in large ROIs than in smaller ones \citep{Liu_2024_MIA_WeReally}. This phenomenon is evident in the examples of poor parcellation, as shown in Fig.~\ref{fig:bad}. In the precentral case, the number of mislabeled vertices is greater than in the parsorbitalis case; however, the impact on the ROI Dice is substantially smaller for the precentral case (a decrease to $44.5\%$ compared to an average of $\sim 90\%$) than for the parsorbitalis case (a decrease to $11.9\%$). Therefore, it is critical to ensure a consistent evaluation framework, where the standard Dice coefficient is used (rather than any normalized or modified versions), and to also assess performance using metrics that are not sensitive to the size of the ROIs. In this study, we also reported point-wise accuracy as a secondary measure.

Another potential source of bias related to the Dice metric arises from the resolution of the input feature maps and the output parcellation maps. When images are downsampled to a tessellated surface with lower resolution, the accuracy of parcellation along the parcel boundaries becomes less relevant, which may artificially inflate the Dice score when evaluation is conducted at the lower resolution. In contrast to methods explicitly trained on lower-resolution surfaces (see Table~\ref{tab:comparison}), JParc operates on a $512\times 256$ grid of parameterization, which maintains a sampling density similar to the subject's native surfaces.

Finally, due to the limited number of subjects available in the Mindboggle dataset, it is essential to perform a full cross-validation to ensure that each subject is tested exactly once for a fair comparison. As with many other approaches \citep{Lombaert_2015_MICaCI-M2_SpectralForests,Gopinath_2019_MIA_GraphConvolutions,Gopinath_2023_MIA_LearningJoint,Parvathaneni_2019_MICCAI-M2_CorticalSurface,Ha_2022_ITMI_SPHARMNetSpherical}, we performed five 5-fold cross-validation in this study and did not compare our results with methods that employed partial cross-validation or a single train-validation-test split.

\subsection{Limitation and Future Directions}
\label{sec:discuss:limit}

One of the limitations of JParc is its reliance on surface reconstruction and the pre-computation of cortical features (e.g., sulcal depth, mean curvature) using software such as FreeSurfer \citep{Fischl_2012_N_FreeSurfer}, which is a time-consuming process. However, with the recent development of several deep learning-based surface reconstruction algorithms, such as Topofit \citep{Hoopes_2022_P5ICMIDL_TopoFitRapid}, this dependency on surface-based preprocessing will no longer be a bottleneck in the near future, and indeed a number of classical algorithms have been replaced with faster and more accurate deep-learning-based alternatives in the recently released FreeSurfer 8.0. Future iterations of JParc will aim to integrate emerging deep learning-based surface reconstruction frameworks directly into the processing pipeline, thereby reducing reliance on classical surface-based preprocessing. Another promising direction is a full transition to end-to-end architectures that directly infer cortical parcellations from raw anatomical MRI, bypassing intermediate feature computation steps \citep{Gopinath_2023_MIA_LearningJoint}.

Another limitation of the current approach is overfitting, a common issue for deep learning models, which is particularly pronounced in this case due to the relatively small sample size of 100 subjects \citep{Klein_2012_FN_101Labeled}. The larger models and increased number of parameters used in many state-of-the-art parcellation methods may exacerbate this overfitting issue, reducing the model's generalizability. One potential avenue for improvement would be to manually label additional subjects and curate a larger dataset for both training and validation.

\begin{figure}[tb]
  \centering
  \includegraphics[width=.6\textwidth]{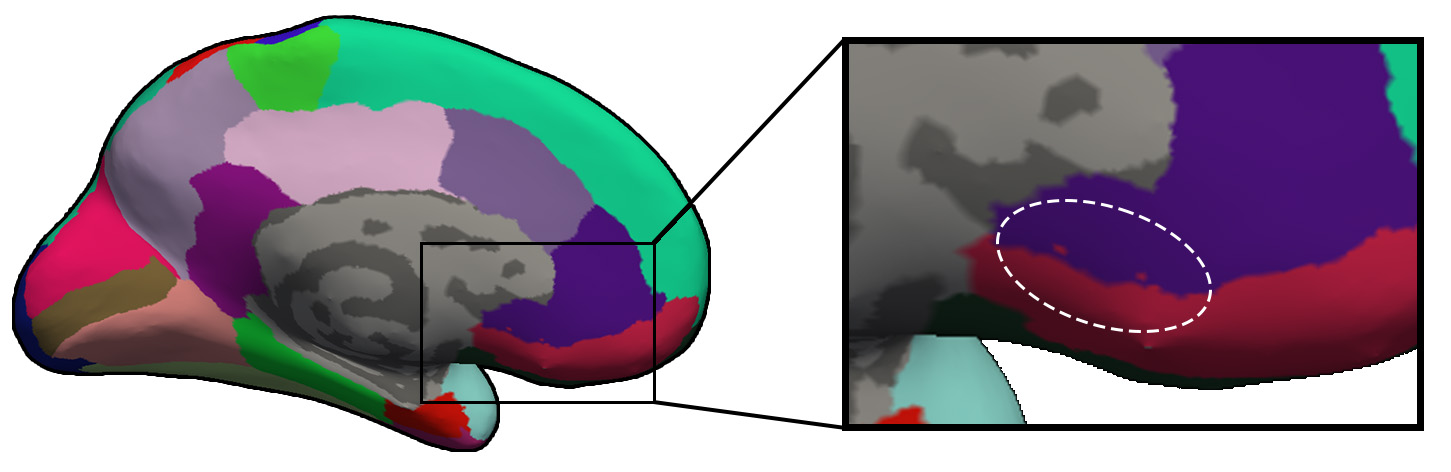}
  \caption{\textbf{An example of JParc parcellation with disjoint labels for the same ROI.} The overall medial surface view is shown on the left and a zoom-in view centered on the medial prefrontal cortex is shown on the right.}
  \label{fig:discontinuity}
\end{figure}

Furthermore, our current model does not impose spatial constraints, which means that parcels may not be spatially contiguous (spatial contiguity is implicitly encouraged in the JOSA registration module due to diffeomorphic deformation fields \citep{Li_2024_MIA_JOSAJoint}). Fig.~\ref{fig:discontinuity} illustrates an example where disjoint labels for the same ROI could be produced, although this happened very infrequently and only in small regions. To address this issue, integrating explicit spatial-contiguity constraints within the network architecture may further improve the anatomical plausibility of parcellations.

Beyond architectural refinements, domain-adaptation strategies could be developed to ensure that JParc maintains high performance across scanners, acquisition protocols, and population cohorts. Finally, incorporating uncertainty estimation into the parcellation outputs could provide quantitative confidence measures for each ROI, thereby improving the reliability of downstream statistical and clinical analyses.

\section*{Data and Code Availability}

The Mindboggle dataset used in this study is publicly available (\url{https://mindboggle.info/data}). JParc code and utility will be integrated into FreeSurfer and available in a future release (\url{https://surfer.nmr.mgh.harvard.edu}).

\section*{Acknowledgments}

Support for this research was provided in part by the BRAIN Initiative Cell Atlas Network (BICAN) grants U01MH117023, UM1MH134812 and UM1MH130981, the Brain Initiative Brain Connects consortium (U01NS132181, 1UM1NS132358), the National Institute for Biomedical Imaging and Bioengineering (1R01EB023281, R21EB018907, R01EB019956, P41EB030006, 1R01EB033773), the National Institute on Aging (R21AG082082, 1R01AG064027, R01AG016495, 1R01AG070988), the National Institute of Mental Health (UM1MH130981, R01MH123195, R01MH121885, 1RF1MH123195), the National Institute for Neurological Disorders and Stroke, (1U24NS135561, R01NS070963, 2R01NS083534, R01NS105820, R25NS125599, R01NS138257, R01NS128961, RF1NS115268, U01NS137484), and was made possible by the resources provided by Shared Instrumentation Grants 1S10RR023401, 1S10RR019307, and 1S10RR023043. Additional support was provided by the NIH Blueprint for Neuroscience Research (5U01-MH093765), part of the multi-institutional Human Connectome Project. Much of the computational resources required for this research were generously provided by the Massachusetts Life Sciences Center (https://www.masslifesciences.com). 

BF and AVD are advisors to DeepHealth, a company whose medical pursuits focus on medical imaging and measurement technologies. AVD is also a consultant for Radence. Both BF's and AVD's interests were reviewed and are managed by Massachusetts General Hospital and Mass General Brigham in accordance with their conflict of interest policies.

\bibliographystyle{model2-names.bst}
\bibliography{refs}

\end{document}